\documentclass[twocolumn,showpacs,aps,prl,letterpaper]{revtex4}
\usepackage{graphicx}
\usepackage{dcolumn}
\usepackage{amsmath}
\usepackage{epsfig}
\usepackage{eepic}
\usepackage{color}

\long\def\inst#1{\par\nobreak\kern 4pt\nobreak
    {\itshape #1}\par\vskip 10pt plus 3pt minus 3pt}
\RequirePackage{xspace}

\def\babar{\mbox{\slshape B\kern-0.1em{\smaller A}\kern-0.1em
    B\kern-0.1em{\smaller A\kern-0.2em R}}}
\def\Kbar    {\kern 0.18em\overline{\kern -0.18em K}{}\xspace}

\def\Kz      {\ensuremath{K^0}\xspace}
\def\Kzb     {\ensuremath{\Kbar^0}\xspace}
\def\KzKzb   {\ensuremath{\Kz {\kern -0.16em \Kzb}}\xspace}

\def\Ks     {\ensuremath{K_S}\xspace}
\def\Kl     {\ensuremath{K_L}\xspace}
\def\KsKs   {\ensuremath{\Ks {\kern -0.16em \Ks}}\xspace}
\def\KlKl   {\ensuremath{\Kl {\kern -0.16em \Kl}}\xspace}
\def\KsKl   {\ensuremath{\Ks {\kern -0.16em \Kl}}\xspace}
\def\KlKs   {\ensuremath{\Kl {\kern -0.16em \Ks}}\xspace}
\def\Dbar    {\kern 0.18em\overline{\kern -0.18em D}{}\xspace}
\def\Dz      {\ensuremath{D^0}\xspace}
\def\Dzb     {\ensuremath{\Dbar^0}\xspace}
\def\DzDzb   {\ensuremath{\Dz {\kern -0.16em \Dzb}}\xspace}
\def\Lam      {\ensuremath{\Lambda}\xspace}
\def\Lamb      {\ensuremath{\overline{\Lambda}}\xspace}
\def\LLb   {\ensuremath{\Lam {\kern -0.16em \Lamb}}\xspace}

\newcommand{\DsP}{\ensuremath{D_s^+}\xspace}
\newcommand{\DsM}{\ensuremath{D_s^-}\xspace}
\newcommand{\DspDsm}{\ensuremath{\DsP {\kern -0.16em \DsM}}\xspace}
\newcommand{\Dp}{\ensuremath{D^+}\xspace}
\newcommand{\Dm}{\ensuremath{D^-}\xspace}
\newcommand{\DpDm}{\ensuremath{\Dp {\kern -0.16em \Dm}}\xspace}

\def\Bbar    {\kern 0.18em\overline{\kern -0.18em B}{}\xspace}

\def\Bz      {\ensuremath{B^0}\xspace}
\def\Bzb     {\ensuremath{\Bbar^0}\xspace}
\def\BzBzb   {\ensuremath{\Bz {\kern -0.16em \Bzb}}\xspace}
\def\Bu      {\ensuremath{B^+}\xspace}
\def\Bub     {\ensuremath{B^-}\xspace}

\def\BpBm    {\ensuremath{\Bu {\kern -0.16em \Bub}}\xspace}

\def\Dp      {\ensuremath{D^+}\xspace}

\newcommand{\optbar}[1]{\shortstack{{\tiny (\rule[.4ex]{1em}{.1mm})}
  \\ [-.7ex] $#1$}}
\def\BorBbar    {\kern 0.18em\optbar{\kern -0.18em B}{}\xspace}
\def\DorDbar    {\kern 0.18em\optbar{\kern -0.18em D}{}\xspace}
\def\KorKbar    {\kern 0.18em\optbar{\kern -0.18em K}{}\xspace}

\def\pep2{PEP-II}
\mathchardef\Upsilon="7107
\def\Y#1S{\ensuremath{\Upsilon{(#1S)}}\xspace}

\begin{document}

\title{\large \bfseries \boldmath Study of $\Lam - \Lamb$ Oscillation in quantum coherent $\Lam\Lamb$ state by using  J/$\psi \rightarrow \LLb$ decay }
\author{Xian-Wei Kang$^{1,2}$}\email{kangxw@ihep.ac.cn}
\author{Hai-Bo Li$^1$}\email{lihb@ihep.ac.cn}
\author{Gong-Ru Lu$^2$}
\affiliation{$^1$Institute of High Energy Physics, P.O.Box 918,
Beijing  100049, China\\ $^2$Department of Physics, Henan Normal
University, Xinxiang 453007, China}


\date{\today}


\begin{abstract}
We discuss the possibility of searching for the $\Lam - \Lamb$
oscillations for coherent $\Lam\Lamb$ production in the J/$\psi
\rightarrow \Lam \Lamb$ decay process. The sensitivity of
measurement of $\Lam -\Lamb$ oscillation in the external field at
BES-III experiment is considered. These considerations indicate an
alternative way to probe the $\Delta B =2$ amplitude in addition to
neutron oscillation experiments. Both coherent and time-dependent
information can be used to extract the $\Lam-\Lamb$ oscillation
parameter. With one year's luminosity at BES-III, we can set an
upper limit of $\delta m_{\Lam \Lamb} < 10^{-15}$ MeV at 90\%
confidence level, corresponding to about $10^{-6}$ s of $\Lam-\Lamb$
oscillation time.

\end{abstract}


\maketitle

One of the open questions for fundamental particle physics is
whether baryon number violation  can be found in
nature~\cite{proton1,proton2}, which is key for understanding the
observed matter antimatter asymmetry. There are a few reasons to
believe that baryon number symmetry may not be exact symmetry. This
is because of the three conditions for generating this asymmetry
pointed out originally by Sakharov in 1967~\cite{sakharov}: (a)
existence of $CP$ violation, (b) baryon number violating
interactions, and (c) the presence of out of thermal equilibrium
conditions in the early Universe. If indeed such interactions are
there, the important question is how one can observe them in
experiments. In 1980, it was pointed out by Marshak {\it et
al.}~\cite{marshak} that a crucial test of baryon number violation
is neutron-antineutron ($N-\overline{N}$) oscillation. Thus,
provided that the new gauge structure occurs beyond the standard
model (SM) the mass scale could be the order of $10^2$ to $10^3$
TeV. After this proposal was made, many experiments had been carried
out for searching for $N-\overline{N}$ oscillation~\cite{nn1,nn2}.
The last experiment in the free neutron system at the ILL sets an
upper limit of $8.6\times 10^7$ s (90\% confidence level) on the
oscillation time~\cite{ill}.

 Moreover, as discussed in Ref.~\cite{obser}, recent discoveries of
neutrino oscillations have made $N-\overline{N}$ oscillation to be
quite plausible theoretically if small neutrino masses are to be
understood as a consequence of the seesaw mechanism~\cite{seesaw},
which indicates the existence of $\Delta(B-L)=2$ interactions.
Therefore, it implies the existence of $N-\overline{N}$
oscillation.

 It is worth noting that if $n -\overline{n}$ oscillation exists,
then $\Lam -\Lamb$ oscillation may also take place as firstly
proposed by K.-B.~Luk~\cite{luk}. In fact one can also consider
$\Lam -\Lamb$ oscillation independently. However, until now there
has not been any direct experimental measurement of $\Lambda-\Lamb$
oscillation. In this letter, we would like to consider the
phenomenology of $\Lam- \Lamb$ oscillation for free $\Lam$. We also
consider the effect of an external field on the $\Lam$ baryon, in
particular, the effect of an external magnetic field on the opposite
magnetic moments of $\Lam$ and $\Lamb$. Moreover, we first propose
to search for e $\Lam-\Lamb$ oscillation in the coherent production
in $J/\psi \rightarrow \Lam\Lamb$ decay. We discuss the observable
for both the time-dependent and the time-independent correlated
production rate.



The time evolution of the $\Lam -\Lamb$ oscillation is described
by the Schr\"odinger-like equation as
\begin{equation}
i \frac{\partial}{\partial t}\,\left(
\begin{array}{c}
\Lam(t)\\
\Lamb(t)
\end{array}\right)
={\bf M} \,\left(
\begin{array}{c}
\Lam(t)\\
\Lamb(t)
\end{array}\right),
\label{eq:Schrodinger}
\end{equation}
where the ${\bf M}$ matrix is Hermitian, and is defined as
\begin{equation}
{\bf M} = \left(
\begin{array}{cc}
m_{\Lam} - \Delta E_{\Lam} & \delta m_{\Lam \Lamb}\\
\delta m_{\Lam \Lamb} & m_{\Lamb} - \Delta E_{\Lamb}\\
\end{array}\right),
\label{eq:operator}
\end{equation}
where $\delta m_{\Lam \Lamb}$ is the $\Delta B =2$ transition mass
between $\Lam$ and $\Lamb$; $m_{\Lam}$ ($m_{\Lamb}$) is the mass of
the $\Lam$ ($\Lamb$) baryon, and $\Delta E_{\Lam}= -
\vec{\mu}_{\Lam} \cdot \vec{B}$ and $\Delta E_{\Lamb}= -
\vec{\mu}_{\Lamb} \cdot \vec{B}$
 are energy split due to external field $\vec{B}$. Here,
 $\vec{\mu}_{\Lam, \Lamb}$ is the magnetic moment of $\Lam$, $\mu_{\Lam} = -\mu_{\Lamb} = - 0.613\mu_N$
($\mu_N=3.152 \times 10^{-14}$ MeV T$^{-1}$ is the nuclear
magneton).
For produced unbound $\Lam$ propagating in a vacuum without an
external field, both $\Delta E_{\Lam}$ and $\Delta E_{\Lamb}$ are
equal to zero. $CPT$ invariance imposes $m_{\Lam} \equiv m_{\Lamb}$
and $\Delta E_{\Lam} = - \Delta E_{\Lamb}$. The equality of the
off-diagonal elements follows from $CP$ invariance. The two
eigenstates $|\Lam_H \rangle$ and $|\Lam_L\rangle$ of the effective
Hamiltonian matrix ${\bf M}$ are given by
\begin{eqnarray}
|\Lam_H \rangle = \frac{1}{\sqrt{2}}(\sqrt{1+z} |\Lam \rangle +
\sqrt{1-z} |\Lamb \rangle ) ,  \nonumber \\
|\Lam_L \rangle = \frac{1}{\sqrt{2}}(\sqrt{1-z} |\Lam \rangle -
\sqrt{1+z}  |\Lamb \rangle )\, , \label{eq:eigenstates}
\end{eqnarray}
where $z = \frac{2 \Delta E}{\Delta m}$. Here, we define $\Delta E =
|\Delta E_{\Lam}| = |\Delta E_{\Lamb}|$ and $\Delta m \equiv m_H-
m_L = 2 \sqrt{(\Delta E)^2 + \delta m^2_{\Lam \Lamb}}$, and $m_H$
[$m_L$] is the mass of the ``heavy (H)"  $\Lam_H$ [``light (L)"
$\Lam_L$] baryon. In the absence of an external field, one has
$\Delta E = 0$, thus one gets $z =0$. While assuming that $\delta
m_{\Lam\Lamb} \sim \delta m_{n\overline{n}}$ for the first order, we
have $\delta m_{\Lam\Lamb} < 10^{-23} $ eV. It indicates that an
external field will make $\Delta m \gg \delta m_{\Lam \Lamb}$, as a
result we have $z \rightarrow 1$.

From Eq.~\eqref{eq:eigenstates}, the corresponding eigenvalues are
\begin{equation}
\lambda_{\Lam_H} = m_{\Lam} + \sqrt{(\Delta E)^2 + \delta
m^2_{\Lam\Lamb}},
 \label{eq:eigenvalue1}
\end{equation}
\begin{equation}
\lambda_{\Lam_L} = m_{\Lam} - \sqrt{(\Delta E)^2 + \delta
 m^2_{\Lam \Lamb} }\, . \label{eq:eigenvalue2}
\end{equation}
 Thus, starting with a beam of $\Lam$, the probability of generating a
$\Lamb$ after time $t$, ${\cal P}(\Lamb, t)$, is described by the
following equation:
\begin{equation}
 {\cal P}(\Lamb, t) = \frac{\delta m^2_{\Lam \Lamb}}{\delta m_{\Lam \Lamb}^2
  + (\Delta E)^2}\mbox {sin}^2(\sqrt{\delta m^2_{\Lam \Lamb} + (\Delta
  E)^2} \cdot t). \label{eq:prob}
\end{equation}
For free $\Lam$, we have $\Delta E =0$, and Eq.~\ref{eq:prob}
becomes
\begin{equation}
 {\cal P}(\Lamb, t) = \mbox{sin}^2(\delta m_{\Lam \Lamb} \cdot t). \label{eq:probfree}
\end{equation}

Hereafter, we consider the possible search of $\Lam -\Lamb$
oscillation in $J/\psi \rightarrow \Lam\Lamb$ decay, in which the
coherent $\Lam \Lamb$ events are generated with a strong boost. Here
we assume that possible strong multiquark effects that involve
seaquarks play no role in $J/\psi \to \Lam \Lamb$
decays~\cite{voloshin}.

In order to satisfy both the angular momentum conservation and
parity conservation, the relative orbital angular momentum of the
$\Lam \Lamb $ pair must be $L=0$ or $2$, and the total spin is $S=1$
({\it i.e.}$^3S_1$ and $^3D_1$ states), then the $\Lam \Lamb $ pair
must be in a state with $C= -1$ \cite{liyang}. Thus, considering
both the spin and orbital part under hypothesis of
``factorization,'' the wave function of the $\Lam \Lamb$ pair system
can be defined as
\begin{equation}\label{wave function}
|\Lam \Lamb \rangle^{C=-1} = \chi_1 \frac{1}{\sqrt{2}} \left [ |\Lam
\rangle |\Lamb\rangle - |\Lamb\rangle |\Lam\rangle\right ].
\end{equation}where $\chi_1$ is the symmetric spin triplet for the fermion pair in the $S=1$ state with $S$ denoting the total spin angular momentum.
Then the amplitude for $J/\psi$ decaying to $\Lam \Lamb$ can be
denoted by $\langle \Lam \Lamb |{\cal H}| J/\psi \rangle$ where
$|\Lam \Lamb\rangle$ is the total wave function for the
$\Lambda\overline\Lambda$ pair. For simplicity, we may write only
the orbital angular part for representing the total wave function
since the occurrence of $\chi_1$ would not affect the genuine
physical results, which can be easily seen in the following
paragraphs.

Now we turn to analyze the time evolution of the $\Lam \Lamb$ system
produced in $J/\psi$ decay. Following $J/\psi \rightarrow \Lam\Lamb$
decay, the $\Lam$ and $\Lamb$ will separate and the proper-time
evolution of the particle states
$|\Lam_{\small\mbox{phys}}(t)\rangle$ and
$|\Lamb_{\small\mbox{phys}}(t)\rangle$ are given by
\begin{eqnarray}
|\Lam_{\small\mbox{phys}}(t) \rangle & = & (g_+(t) + z g_-(t)
|\Lam \rangle + \sqrt{1-z^2} g_-(t) |\Lamb \rangle, \nonumber \\
|\Lamb_{\small\mbox{phys}}(t) \rangle & = & (z g_-(t) -g_+(t))
|\Lamb \rangle - \sqrt{1-z^2} g_-(t) |\Lam \rangle, \nonumber \\
\label{eq:lam_time}
\end{eqnarray}
where
\begin{eqnarray}
g_{\pm} = \frac{1}{2} (e^{-im_H t-\frac{1}{2} \Gamma_H t} \pm
e^{-im_L t - \frac{1}{2}\Gamma_L t} ),
 \label{eq:define}
\end{eqnarray}
with definitions
\begin{eqnarray}
m &\equiv& \frac{m_L + m_H}{2}, \, \, \Delta m \equiv m_H - m_L,
\nonumber \\
\Gamma &\equiv & \frac{\Gamma_L + \Gamma_H}{2}, \, \Delta \Gamma
\equiv \Gamma_H - \Gamma_L,
 \label{eq:define2}
\end{eqnarray}
Note that here, $\Delta m$ is positive by definition, while the
sign of $\Delta \Gamma$ is to be determined by experiments.

In practice, one defines the following oscillation parameters in a
similar fashion as in neutral $B$ and $D$ mixing cases:
\begin{eqnarray}
 x_{\Lam} \equiv \frac{\Delta m}{\Gamma}, \, y_{\Lam} \equiv \frac{\Delta
 \Gamma}{2\Gamma}.
\label{eq:define3}
\end{eqnarray}
Then we consider a $\Lam \Lamb$ pair in $J/\psi$ decay with a
definite charge-conjugation eigenvalue. The time-dependent wave
function of $\Lam \Lamb$ system with $C=-1$ can be written as
\begin{eqnarray}
|\Lam\Lamb (t_1,t_2) \rangle &=& \frac{1}{\sqrt{2}} [
|\Lam_{\small\mbox{phys}}({\bf k_1},t_1)
\rangle|\Lamb_{\small\mbox{phys}}({\bf k_2}, t_2) \rangle \nonumber \\
 & - & |\Lamb_{\small\mbox{phys}}({\bf k_1}, t_1) \rangle |\Lam_{\small\mbox{phys}}({\bf k_2}, t_2) \rangle ],
 \label{eq:lamlamb_time}
\end{eqnarray}
where ${\bf k_1}$ and ${\bf k_2}$ are the three-momentum vector of
the two $\Lam$ baryons.  We now consider decays of these
correlated system into various final states. The amplitude of such
joint decays, one $\Lam$ decaying to a final state $f_1$ at proper
time $t_1$, and the other $\Lam$ to $f_2$ at proper time $t_2$, is
given by
\begin{eqnarray}
A(J/\psi &\to& \Lam_{\small\mbox{phys}}\Lamb_{\small\mbox{phys}}
\to f_1 f_2) \equiv \frac{1}{\sqrt{2}}\times \nonumber \\
&& \{ [g_+(t_1)g_-(t_2) - g_-(t_1)g_+(t_2)] a_2 - \nonumber \\
&& [g_+(t_1)g_+(t_2)-g_-(t_1)g_-(t_2) ] a_1\},
 \label{eq:amp}
\end{eqnarray}
where
\begin{eqnarray}
a_1 &\equiv & A_{f_1}\overline{A}_{f_2} -
\overline{A}_{f_1}A_{f_2} = A_{f_1}A_{f_2} (\lambda_{f_2} -
\lambda_{f_1}),
\nonumber \\
 a_2 &\equiv& z(A_{f_1} \overline{A}_{f_2} +
\overline{A}_{f_1} A_{f_2}) - \sqrt{1-z^2}(A_{f_1} A_{f_2} -
\overline{A}_{f_1}\overline{A}_{f_2}) \nonumber \\
 &=& A_{f_1}A_{f_2} [ z(\lambda_{f_2} + \lambda_{f_1}) -
 \sqrt{1-z^2}(1-\lambda_{f_1}\lambda_{f_2})],
 \label{eq:aa}
\end{eqnarray}
with $A_{f_i} \equiv \langle f_i|{\cal H}|\Lam\rangle$,
$\overline{A}_{f_i} \equiv \langle f_i|{\cal H}|\Lamb\rangle$ ($i
=1,2$), and we define
\begin{eqnarray}
\lambda_{f_i} \equiv
 \frac{\langle f_i | {\cal H}|\Lamb\rangle}{\langle f_i|{\cal
 H}|\Lam\rangle} =\frac{\overline{A}_{f_i}}{A_{f_i}},
\label{eq:lambda_define}
\end{eqnarray}
\begin{eqnarray}
\overline{\lambda}_{\overline{f}_i} \equiv
 \frac{\langle \overline{f_i} | {\cal H}|\Lam\rangle}{\langle \overline{f_i}|{\cal
 H}|\Lamb\rangle}
 =\frac{A_{\overline{f_i}}}{\overline{A}_{\overline{f_i}}}.
\label{eq:lambda_define2}
\end{eqnarray}

 In the process $e^+e^- \to J/\psi \to \Lam\Lamb$, the $\Lam\Lamb$ pairs are strongly boosted,
 so that the decay-time difference [t=$\Delta
t_- = (t_2 - t_1)$] between $\Lam_{\small\mbox{phys}} \to f_1$ and
$\Lamb_{\small\mbox{phys}} \to f_2$ can be measured easily. From
Eq.~(\ref{eq:amp}),  one can derive the general expression for the
time-dependent decay rate:
\begin{eqnarray}
&&\frac{d\Gamma(J/\psi\to
\Lam_{\small\mbox{phys}}\Lamb_{\small\mbox{phys}} \to f_1
f_2)}{dt} =  {\cal N} e^{-\Gamma|t|}\times\nonumber \\
&&[(|a_1|^2+|a_2|^2)\mbox{cosh}(y_{\Lambda}\Gamma t) + (|a_1|^2 -
|a_2|^2)\mbox{cos}(x_{\Lambda}\Gamma t) \nonumber \\
&& +2{\cal R}e(a_1 a^*_2)\mbox{sinh}(y_{\Lambda}\Gamma t) +2{\cal
I}m(a_1 a^*_2)\mbox{sin}(x_{\Lambda}\Gamma t)],
 \label{eq:decay_rate}
\end{eqnarray}
where ${\cal N}$ is a common normalization factor. In
Eq.~(\ref{eq:decay_rate}), terms proportional to $|a_1|^2$ are
associated with decays that occur without any net oscillation, while
terms proportional to $|a_2|^2$ are associated with decays following
a net oscillation. The other terms are associated with the
interference between these two cases. In the following discussion,
we define
\begin{eqnarray}
R(f_1,f_2; t) \equiv \frac{d\Gamma(J/\psi \to
\Lam_{\small\mbox{phys}}\Lamb_{\small\mbox{phys}} \to f_1
f_2)}{dt}.
 \label{eq:decay_rate_define}
\end{eqnarray}

 For a given state $f_1 f_2 = (p\pi^-)
(p\pi^-)$, we have $a_1 = 0$ and $a_2 = 2 A_{p\pi^-}
\overline{A}_{p\pi^-}$. Thus one can write $R(p\pi^-,p\pi^-; t)$
as:
\begin{eqnarray}
  R(p\pi^-,p\pi^-; t) = {\cal N}
\frac{1}{4} e^{-\Gamma| t|}|a_2|^2 [\mbox{cosh}(y_{\Lambda}\Gamma
t)
  -\mbox{cos}(x_{\Lambda}\Gamma t)]. \nonumber \\
 \label{eq:decay_ratesim}
\end{eqnarray}
At BES-III experiment, the external magnetic field is about 1.0 T,
in which case, $\Delta E \sim 2\times 10^{-11}$ MeV, thus $z \sim
1$.  Taking into account that $|\lambda|$, $|\overline{\lambda}| \ll
1$ and $x_{\Lam}$, $y_{\Lam} \ll 1$ and $z \rightarrow 1.0$, keeping
terms up to order $x_{\Lam}^2$, and $y^2_{\Lam}$ in the expressions,
neglecting $CP$ violation, expanding the time-dependent for $x t$,
$y t$,  we can write Eq.~ \eqref{eq:decay_ratesim} as
\begin{eqnarray}
  R(p\pi^-,p\pi^-; t) = {\cal N}
 e^{-\Gamma| t|} |A_{p\pi^-}|^2 |\overline{A}_{p\pi^-}|^2
 \frac{x_{\Lam}^2 +y^2_{\Lam}}{2} (\Gamma t)^2. \nonumber \\
 \label{eq:decay_rate1}
\end{eqnarray}

For $f_1 f_2 =(p\pi^-)(\overline{p}\pi^+)$, we have $a_1 =
A_{p\pi^-}\overline{A}_{\overline{p}\pi^+}
(1-\lambda_{p\pi^-}\overline{\lambda}_{\overline{p}\pi^+})$ and $a_2
= A_{p\pi^-}\overline{A}_{\overline{p}\pi^+}
(1+\lambda_{p\pi^-}\overline{\lambda}_{\overline{p}\pi^+})$. Thus
the time-dependent decay rate can be expressed as
\begin{eqnarray}
  R(p\pi^-,\overline{p}\pi^+; t) &=& {\cal N} \frac{1}{2}
 e^{-\Gamma| t|} |A_{p\pi^-}|^2 |\overline{A}_{p\pi^-}|^2
 (1+ y_{\Lambda} \Gamma t) \nonumber \\
 &\approx &{\cal N} \frac{1}{2}
 e^{-\Gamma| t|} |A_{p\pi^-}|^2 |\overline{A}_{p\pi^-}|^2.
 \label{eq:decay_rate2}
\end{eqnarray}

We define the following observable :
\begin{eqnarray}\label{ratio}
{\cal R}(t) \equiv  \frac{ R(p\pi^-,p\pi^-; t) +
R(\overline{p}\pi^+,\overline{p}\pi^+;
t)}{R(p\pi^-,\overline{p}\pi^+; t)+ R(\overline{p}\pi^+,p\pi^-;
t)}
 \label{eq:cptime}
\end{eqnarray}
Combining Eqs.~\eqref{eq:decay_rate1} and~\eqref{eq:decay_rate2},
 one obtains:
\begin{eqnarray}
{\cal R}(t) =  2|\lambda_{p\pi^-}|^2 \frac{x^2_{\Lam}
+y^2_{\Lam}}{2} (\Gamma t)^2,
 \label{eq:cptime-z}
\end{eqnarray}

For completeness, we derive general expressions for
time-integrated decay rates into a pair of final states $f_1$ and
$f_2$:
\begin{eqnarray}
 R(f_1,f_2)  &= &\frac{1}{4}{\cal N}
 \big[(|a_1|^2+|a_2|^2)\frac{1}{1-y^2_{\Lambda}}\nonumber \\
    && +(|a_1|^2-|a_2|^2)\frac{1}{1+x^2_{\Lambda}}\big].
 \label{eq:cp_independ}
\end{eqnarray}
At last the ratio of two probabilities mentioned above can be
rewritten as
\begin{equation}
{\cal R} \equiv \frac{ R(p\pi^-,p\pi^-) +
R(\overline{p}\pi^+,\overline{p}\pi^+)}{R(p\pi^-,\overline{p}\pi^+)+
R(\overline{p}\pi^+,p\pi^-)} = 2 |\lambda_{p\pi^-}|^2
(x_{\Lam}^2+y_{\Lam}^2) . \label{eq:inter}
\end{equation}

If there is no external field and the $\Lam$ is free, we have $z=0$,
Eq.~\eqref{eq:cptime-z} becomes
\begin{eqnarray}
{\cal R}(t) =  \frac{1}{2} \frac{x^2_{\Lam} +y^2_{\Lam}}{2}
(\Gamma t)^2,
 \label{eq:cptimez0}
\end{eqnarray}
and the time-independent ratio in Eq.~(\ref{eq:inter}) becomes
\begin{eqnarray}
{\cal R} =  \frac{x^2_{\Lam} +y^2_{\Lam}}{2}.
 \label{eq:cpintez0}
\end{eqnarray}
Assuming $y_\Lambda=0$, one estimation value of $\delta
m_{\Lambda\bar\Lambda}$ in the presence of an external field reads
from Eq.~\eqref{eq:inter},
\begin{equation}\label{delta field}
\delta
m_{\Lambda\bar\Lambda}=\sqrt{(\frac{\mathcal{R}\Gamma}{4|\lambda_{p\pi^-}|^2})^2-(\Delta
E)^2}
\end{equation} Correspondingly, Eq.~\eqref{eq:cpintez0} is rewritten
as
\begin{equation}\label{delta nofield}
\delta m_{\Lambda\bar{\Lambda}} =
\frac{1}{\sqrt{2}}\sqrt{\mathcal{R}}\Gamma
\end{equation}
With huge data sample, one can measure $\mathcal R$ and
$|\lambda_{p\pi^-}|^2$ simultaneously, and from Eq.(29), the
oscillation mass $\delta m_{\Lambda\bar\Lambda}$ will be
determined..

Currently, we can get an estimated value for $\delta
m_{\Lambda\bar\Lambda}$ in the absence of an external field from
Eq.~\eqref{delta nofield}. In the experiment at BES-III, about
$10\times 10^9$ $J/\psi$ and $3\times 10^{9}$ $\psi(2S)$ data
samples can be collected per year's running according to the
designed luminosity of BEPCII in Beijing~\cite{besiii,bepcii}.
Assuming that no signal events of $J/\psi \rightarrow \Lam_H \Lam_L
\rightarrow ( p\pi^-)( p\pi^-)$ or
$(\overline{p}\pi^+)(\overline{p}\pi^+)$ are observed, we can set an
upper limit of $\mathcal{R}<3.5\times10^{-7}$ and further $\delta m_
{\Lam \Lamb}< 10^{-15}$ MeV at 90\% confidence level. This will be
the first search for $\Lam -\Lamb$ oscillation experimentally. In
the future, at the next generation of a $\tau$-charm factory with
luminosity of $10^{35}$ cm$^{-2}$s$^{-1}$
\cite{taucharm1,taucharm2}, the expected sensitivity of measurement
of $\Lam-\Lamb$ oscillation would be more stringent, $\delta m_{\Lam
\Lamb} < 10^{-17}$ MeV at 90\% confidence level.

It is known that one has to fit the proper-time distribution as
described in Eq.~\eqref{eq:cptime-z} in experiments to extract the
$\Lam$ oscillation parameters.  At a symmetric $\psi$ factory,
namely, the $J/\psi$ is at rest in the central-mass (CM) frame.
Then, the proper-time interval between the two $\Lam$ baryons is
calculated as
\begin{eqnarray}
  \Delta t = (r_{\Lam} -r_{\Lamb})\frac{m_{\Lam}}{c{\bf |P|}},
 \label{eq:propertime}
\end{eqnarray}
where $r_{\Lam}$ and $r_{\Lamb}$ are the $\Lam$ and the $\Lamb$
decay lengths, respectively, and ${\bf P}$ is the three-momentum
vector of $\Lam$. Since the momentum can be calculated with $J/\psi$
decay in the CM frame, all the joint $\Lam\Lamb$ decays in this
paper can be used to study $\Lam-\Lamb$ oscillation  in the
symmetric $J/\psi$ factory.

The average decay length of the $\Lam$ baryon in the rest frame of
$J/\psi $ is $c\tau_{\Lam} \times (\beta \gamma)_{\Lam} \approx 7.6$
cm. At BES-III,  the impact parameter resolution of the main draft
chamber, which is directly related to the decay vertex resolution of
$\Lam$, is described in Ref.~\cite{besiii}, from which we can get
that the resolution for the reconstructed $\Lam$ decay length should
be less than $200 \mu$m within the coverage of the detector. This
means that the BES-III detector is good enough to separate the two
$\Lam$ decay vertices, so that the oscillation parameters can be
measured by using time information.

In conclusion, if $n-\overline{n}$ oscillation exists, then it would
be possible to induce $\Lam-\Lamb$ oscillation. We suggest that the
coherent $\Lam\Lamb$ events from the decay of $J/\psi \rightarrow
\Lam\Lamb$ can be used to search for possible $\Lam -\Lamb$
oscillation. The $\Lam$ baryons from $J/\psi$ decay are strongly
boosted, so that it will offer the possibility to measure the
proper-time interval $\Delta t$ between the fully reconstructed
$\Lam$ and $\Lamb$. Both coherent and time-dependent information can
be used to extract the $\Lam-\Lamb$ oscillation parameter. With one
year's luminosity at BES-III, we can set an upper limit of $\delta
m_{\Lam \Lamb} < 10^{-15}$ MeV at 90\% confidence level,
corresponding to about $10^{-6}$ s of $\Lam-\Lamb$ oscillation time.
It will be the first search of $\Lam -\Lamb$ oscillation
experimentally.  At the BES-III experiment, $\Lambda {\bar \Lambda}$
pair can be fully reconstructed, and backgrounds will be highly
suppressed with particle identification and reconstruction of second
vertex of the $\Lambda$ decay. The BES-III experiment is collecting
data at the $J/\psi$ peak now, and we expect to see the first result
of $\Lam-\Lamb$ oscillation sooner. Here we want to point out that
the shorter mean life of $\Lam$ can significantly hamper a sensitive
search for the $\Lam-\Lamb$ oscillation as stated in Ref~\cite{luk}.
Finally, we have to address that the future super $\tau$-charm
factory will be important to search for this kind of new physics.
Precisely measuring the baryon number violating process is
encouraged.


The authors would like to thank M.~Z.~Yang for useful discussions.
This work is supported in part by the National Natural Science
Foundation of China under Contracts No. 10521003 and No. 10821063,
the 100 Talents program of CAS, and the Knowledge Innovation Project
of CAS under Contracts No. U-612 and No. U-530 (IHEP).





\begin{thebibliography}{99}

\bibitem{proton1} H.~Georgi and S.~L.~Glashow, Phy. Rev. Lett.
{\bf 32}, 438 (1974).
\bibitem{proton2} S.~Dimopoulous, S.~Raby, and F.~Wilczek, Phys.
Lett. {\bf 112B}, 133 (1982).
\bibitem{sakharov} A.~D.~Sakharov, JETP Lett. {\bf 5}, 24 (1967).
\bibitem{marshak} R.~N.~Mohapatra and R.~E.~Marshak, Phys. Rev.
Lett. {\bf 44}, 1316 (1980).
\bibitem{nn1}V.~A.Kuzmin, Pis'ma Zh.~Eksp.~Teor.~Fiz. {\bf 12} 335 (1970).

\bibitem{nn2} R.~N.~Mohapatra and R.~E.~Marshak, Phys. Rev. Lett.
{\bf 44}, 1316 (1980).

\bibitem{ill} M.~Baldo-Ceolin {\it et al.}, Z.~Phys.~C {\bf 63},
409 (1994).

\bibitem{obser} B.~Dutta, Y.~Mimura, and R.~N.~Mohapatra, Phys.
Rev. Lett. {\bf 96}, 061801 (2006).

\bibitem{seesaw} P.~Minkowski, Phys. Lett. {\bf 67B}, 421 (1977);
R.~N.~Mohapatra and G.~Senjanovic, Phys. Rev. Lett. {\bf 44}, 912
(1980).

\bibitem{luk} Kam-Biu Luk, ``on $\Lam-\Lamb$ oscillation'', in International Workshop
 on the Seach for Baryon and Lepton Number Violations, Lawrence Berkeley National Laboratory,
Berkeley, California, 2007.

\bibitem{voloshin} M.~B.~Voloshin, Phys.~Rev.~D {\bf 71}, 114003 (2005).
\bibitem{liyang} H.~B.~Li and M.~Z.~Yang, Phys.~Rev.~D {\bf 74},
094016 (2006).

\bibitem{besiii}
BESIII Collaboration, "The Preliminary Design Report of the BESIII
Detector", Report No. IHEP-BEPCII-SB-13.

\bibitem{bepcii} D.~M.~Asner{\it et. al.}, Special issue on Physics at BES-III,
edited by K.~T.~Chao and Y.~F.~Wang [Int.~J.~Mod.~Phys.~A {\bf 24},
5 (2009)].
\bibitem{taucharm1}
 D.~M.~Asner, in {\it``Discoveries in Flavour Physics at $e^+e^-$
Colliders''}, edited by L.~Benussi {\it et al.}, Frascati Physics
Series Vol.~41 (INFN-LNF SIS Ufficio Pubblicazioni, Frascati, Italy,
2006), p.~377.
\bibitem{taucharm2}
 M. Bona {\it et al.}, arxiv:0709.0451.

\end{thebibliography}
\end{document}